\documentclass[fleqn,usenatbib]{mnras}

\usepackage{newtxtext,newtxmath}
\usepackage{subcaption}
\usepackage{enumitem}

\usepackage[T1]{fontenc}

\DeclareRobustCommand{\VAN}[3]{#2}
\let\VANthebibliography\thebibliography
\def\thebibliography{\DeclareRobustCommand{\VAN}[3]{##3}\VANthebibliography}

\usepackage{graphicx}	
\usepackage{amsmath}	
\usepackage{caption}
\usepackage{fancyhdr}
\usepackage{hyperref}
\usepackage{makecell}
\usepackage[table]{xcolor}
\usepackage{colortbl}
\usepackage{enumitem} 
\usepackage{comment}

\graphicspath{ {./figures/} }

\definecolor{darkgreen}{rgb}{0.0,0.65,0.0}

\title[Radio Detection of the Be/X-ray Binary A0538–66]{Detection of an Extremely Luminous Radio Counterpart to the Be/X-ray Binary A0538–66}

\author[J. Crook-Mansour et al.]{
Justine Crook-Mansour$^{1}$\thanks{E-mail: justine.crook-mansour@physics.ox.ac.uk},
Rob Fender$^{1,2}$, Alex Andersson$^{1}$, Hao Qiu$^{3}$, Andrew K. Hughes$^{1}$, \newauthor
Jakob van den Eijnden$^{4}$, Fraser J. Cowie$^{1}$, Sara Motta$^{1,5}$, Itumeleng Monageng$^{2,6}$, Lorenzo Ducci$^{7}$, \newauthor
Sandro Mereghetti$^{8}$, Andries Mathiba$^{2,6}$, Dougal Dobie$^{9,10}$, Tara Murphy$^{9,10}$, David L.~Kaplan$^{11}$, \newauthor 
Francesco Carotenuto$^{12}$, Phil Charles$^{1,13,14}$
\\
$^{1}$Astrophysics, Department of Physics, University of Oxford, Keble Road, Oxford, OX1 3RH, UK\\
$^{2}$Department of Astronomy, University of Cape Town, Private Bag X3, Rondebosch 7701, South Africa\\
$^{3}$SKA Observatory, Jodrell Bank, Lower Withington, Macclesfield SK11 9FT, UK \\
$^{4}$Anton Pannekoek Institute for Astronomy, Universiteit van Amsterdam, Science Park 904, NL-1098 XH Amsterdam, the Netherlands \\
$^{5}$ Istituto Nazionale di Astrofisica (INAF), Osservatorio Astronomico di Brera, via E. Bianchi 46, 23807 Merate (LC), Italy\\
$^{6}$South African Astronomical Observatory, PO Box 9, Observatory, Cape Town 7935, South Africa \\
$^{7}$Institut für Astronomie und Astrophysik, Kepler Center for Astro and Particle Physics, Universität Tübingen, Sand 1, 72076 Tübingen, Germany \\
$^{8}$INAF – Istituto di Astrofisica Spaziale e Fisica Cosmica, Via A. Corti 12, 20133 Milano, Italy \\
$^{9}$Sydney Institute for Astronomy, School of Physics, The University of Sydney, Sydney, 2006, NSW, Australia\\
$^{10}$ARC Centre of Excellence for Gravitational Wave Discovery (OzGrav), Hawthorn, 3122, Victoria, Australia\\
$^{11}$Center for Gravitation, Cosmology, and Astrophysics, Department of Physics \& Astronomy, University of Wisconsin-Milwaukee, P.O. Box 413,\\ Milwaukee, 53201, WI, USA \\
$^{12}$INAF, Osservatorio Astronomico di Roma, Via Frascati 33, I-00078 Monte Porzio Catone, Italy \\
$^{13}$Department of Physics \& Astronomy, University of Southampton, Southampton SO17 1BJ, UK \\
$^{14}$Department of Physics, University of the Free State, PO Box 339, Bloemfontein 9300, South Africa
}

\date{Accepted XXX. Received YYY; in original form ZZZ}

\pubyear{\the\year{}}

\begin{document}
\label{firstpage}
\pagerange{\pageref{firstpage}--\pageref{lastpage}}
\maketitle

\begin{abstract}
We present the discovery of radio emission from the Be/X-ray binary A0538–66 with the Australian Square Kilometre Array Pathfinder (ASKAP), and results from a subsequent weekly monitoring campaign with the MeerKAT radio telescope. A0538–66, located in the Large Magellanic Cloud, hosts a neutron star with a short spin period ($P \approx 69$ ms) in a highly eccentric $\approx16.6$-day orbit. Its rare episodes of super-Eddington accretion, rapid optical and X-ray flares, and other peculiar properties make it an interesting system among high-mass X-ray binaries. Our MeerKAT data reveal that it is also one of the most radio-luminous neutron star X-ray binaries observed to date, reaching $\approx 3 \times 10^{22}~\text{erg}~\text{s}^{-1} \text{Hz}^{-1}$, with radio emission that appears to be orbitally modulated. We consider several possible mechanisms for the radio emission, and place A0538–66 in context by comparing it to similar systems. 
\end{abstract}

\begin{keywords}
accretion -- accretion discs -- stars: neutron -- X-rays: binaries -- X-rays: individual: 1A 0538–66
\end{keywords}





\section{Introduction}

X-ray binaries (XRBs) are bright X-ray sources in which a compact object -- either a neutron star (NS) or black hole (BH) -- accretes matter from a stellar binary companion (donor star). \emph{Be/X-ray binaries (Be/XRBs}; see \citealt{reig_2011} for a review) are a type of high-mass XRB (HMXB) hosting a NS accreting material from an early-type Oe/Be star of luminosity class III–V. The emission lines in the donor star's optical spectra (most commonly studied are the Balmer and Paschen series) and enhanced infrared radiation indicate the presence of a disc of dense gas lying in the equatorial plane of the donor star, known as the \emph{circumstellar/decretion disc}. Accretion-driven outbursts are classically divided into two types: \emph{Type~I} outbursts\footnote{Not to be confused with `Type I X-ray bursts' in low-mass XRBs (LMXBs).} occur when the NS intersects the Be star’s decretion disc near periastron, leading to episodic accretion and typical peaks of $L_X \lesssim 10^{37}\,\mathrm{erg\,s^{-1}}$ (e.g., \citealt{okazaki_2001}), while \emph{Type~II} (giant) outbursts may last multiple orbital cycles and can reach (super-)Eddington luminosities (e.g., \citealt{moritani_2013}).

A0538–66 (also known as 1A 0535–668) is a Be/XRB in the Large Magellanic Cloud (LMC; $D\approx50$ kpc; \citealt{pietrzynski_2019}) that was discovered in 1977 when the Ariel V satellite detected two X-ray outbursts separated by $\sim$17 days \citep{white_1978}. The system contains a NS in a highly eccentric orbit ($e \approx$ 0.72), with a period of $P_{\rm orb}\approx16.6$ days around a B1 IIIe donor \citep{rajoelimanana_2017}. The inclination of the binary plane to our line of sight is constrained to be high (but $\lesssim 75^\circ$, owing to the absence of X-ray eclipses), while the decretion disc is inferred to be misaligned with respect to this and may precess \citep{rajoelimanana_2017, martin_2024}.

The source undergoes X-ray outbursts occurring on a range of timescales, with luminosities spanning five orders of magnitude in dynamical range. In the years following its discovery, several X-ray events were reported to exceed the isotropic Eddington limit (for a NS, $L_{\text{Edd}} \sim 10^{38}~\text{erg~s}^{-1}$; e.g., \citealt{white_1978}), whereas later observations found lower luminosities ($\sim$$10^{33-37}$ erg s$^{-1}$; e.g., \citealt{campana_2002}, \citealt{kretschmar_2004})\footnote{Some super-Eddington outbursts may have been missed or confused in all-sky survey data because A0538–66 lies only $\sim$30 arcmin from the luminous LMC X–4, and its bright episodes are typically short.}. However, in 2018, the source once again reached Eddington levels (with a peak 0.2–10 keV luminosity of $\sim$$4 \times 10^{38}$ erg s$^{-1}$; \citealt{ducci_2019c}). During this time, it exhibited rapid, strong X-ray flares near periastron -- with flux changes of up to three orders of magnitude on timescales of a few seconds -- which may be attributed to very fast switches between spherical accretion and supersonic propeller regimes \citep{ducci_2019c,rigoselli_2025}. Occasionally, the source exhibits X-ray flares far from periastron, which could be due to a warped decretion disc (supported by observations from \citealt{rajoelimanana_2017}) or inhomogeneities/clumps in the surrounding circumstellar environment \citep{ducci_2022}. On 2025 September 30, the source again reached super-Eddington levels, with a 0.2–12 keV luminosity of $\sim$$1.5 \times 10^{39}$ erg s$^{-1}$ \citep{ducci_2025_atel}.

A0538–66 displays X-ray pulsations at $\sim$69 ms \citep{skinner_1982, ducci_2025}, making it the fastest spinning accretion-powered NS in a HMXB\footnote{SAX J0635.2+0533 is reported to have a shorter period, but is more likely rotation-powered \citep{la_palombara_2017}.}. Interestingly, these pulsations have only been detected three times: during a super-Eddington outburst in 1980 \citep{skinner_1982}, in a single $\sim$11-minute observation in 2023 January at a much lower luminosity ($L_{\rm X}\sim8\times10^{36}$ erg s$^{-1}$, 0.3–10 keV; \citealt{ducci_2025}), and during the super-Eddington outburst in 2025 September \citep{ducci_2025_atel}. Their sporadic appearance has been attributed to plasma temporarily leaking through the centrifugal barrier via an instability mechanism, allowing accretion onto the polar caps \citep{ducci_2025}.

The source's optical outbursts also vary widely in amplitude and duration, reaching some of the brightest levels seen in HMXBs ($L_{\rm opt}\sim3\times10^{38}$ erg s$^{-1}$; \citealt{skinner_1980}), likely primarily powered by the reprocessing of X-rays in an envelope of dense gas surrounding the donor star \citep{ducci_2019b}. Its optical emission is orbitally-modulated \citep{skinner_1980}, with double-peaked profiles near periastron supporting a disc-plane misalignment (e.g., \citealt{rajoelimanana_2017}, \citealt{ducci_2025}). The source also displays irregular optical variability on $\sim$day timescales, possibly caused by density perturbations in the decretion disc due to tidal interactions with the NS \citep{ducci_2016, rajoelimanana_2017}.

On longer timescales, photometry revealed a $\sim$420-day \emph{super-orbital} optical modulation, which could be caused by the build-up and depletion of a high-inclination decretion disc, which toggles the system between different states of activity \citep{alcock_2001,mcGowan_2003,rajoelimanana_2017}. Under this interpretation, during \emph{active/flaring} phases, the system is optically fainter because the disc partially occults the B star -- consistent with Balmer (and occasional He I) emission lines and reddening -- while accretion as the NS crosses the disc triggers X-ray and optical flares. Successive periastron passages may deplete the disc, producing \emph{quiescent} intervals when the optical emission is relatively stable and at its brightest, dominated by the naked B star. However, weak orbital outbursts sometimes persist, and since about 2010, the quiescent phases have shortened, suggesting that the disc is more persistently present \citep{schmidtke_2014, rajoelimanana_2017, ducci_2016, ducci_2022}. Alternatively, the super-orbital modulation may arise from precession of the decretion disc \citep{martin_2024}.

In Sections \ref{sec: discovery} and \ref{sec: meerkat}, we present the detection of the radio counterpart to A0538–66, and our subsequent weekly radio monitoring campaign. Sections \ref{sec: swift} and \ref{sec: atlas} describe our analysis of quasi-simultaneous X-ray and optical data, respectively. Finally, in Section \ref{sec: discussion}, we discuss possible origins of the radio emission, deferring a more detailed analysis to a forthcoming paper once additional multi-wavelength data has been obtained. 

We define phase zero as the optical maximum, adopting the ephemeris of \cite{ducci_2022} ($P_{\text{orb}} = 16.64002 \pm 0.00026$ days, $T_0 = \text{MJD}~ 55673.71 \pm 0.05$), which aligns better with our optical peak than that of \cite{rajoelimanana_2017}. The phase-folded light curves are centered on $T_0$, such that $-0.5<\phi \leq 0.0$ and $0.0<\phi \leq 0.5$ correspond to the half-orbits preceding and following the optical maxima, respectively. From radial-velocity fits, \cite{rajoelimanana_2017} placed periastron at $\phi_\text{peri}\approx0.04$, i.e. $\sim$0.7 days after the optical maximum, although no revised estimates are available.


\section{Data Analysis}

\subsection{ASKAP Discovery}\label{sec: discovery}

\begin{figure}
  \centering
  \begin{subfigure}[b]{\linewidth}
    \centering
    \includegraphics[width=\columnwidth]{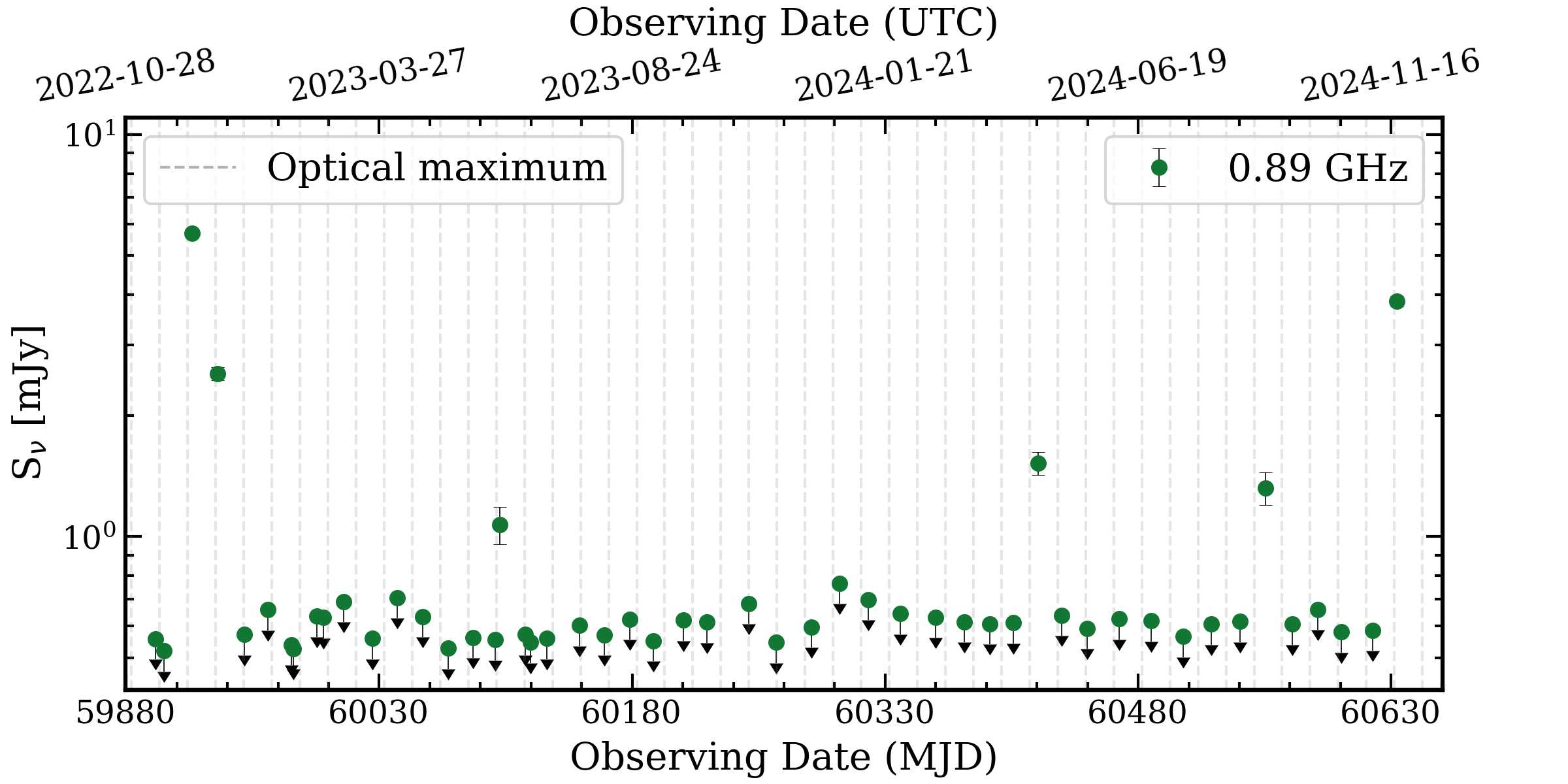}
    \label{fig:vast:lightcurve}
    \vspace{-1.1em}
  \end{subfigure}
  \begin{subfigure}[b]{\linewidth}
    \centering
    \includegraphics[width=\columnwidth]{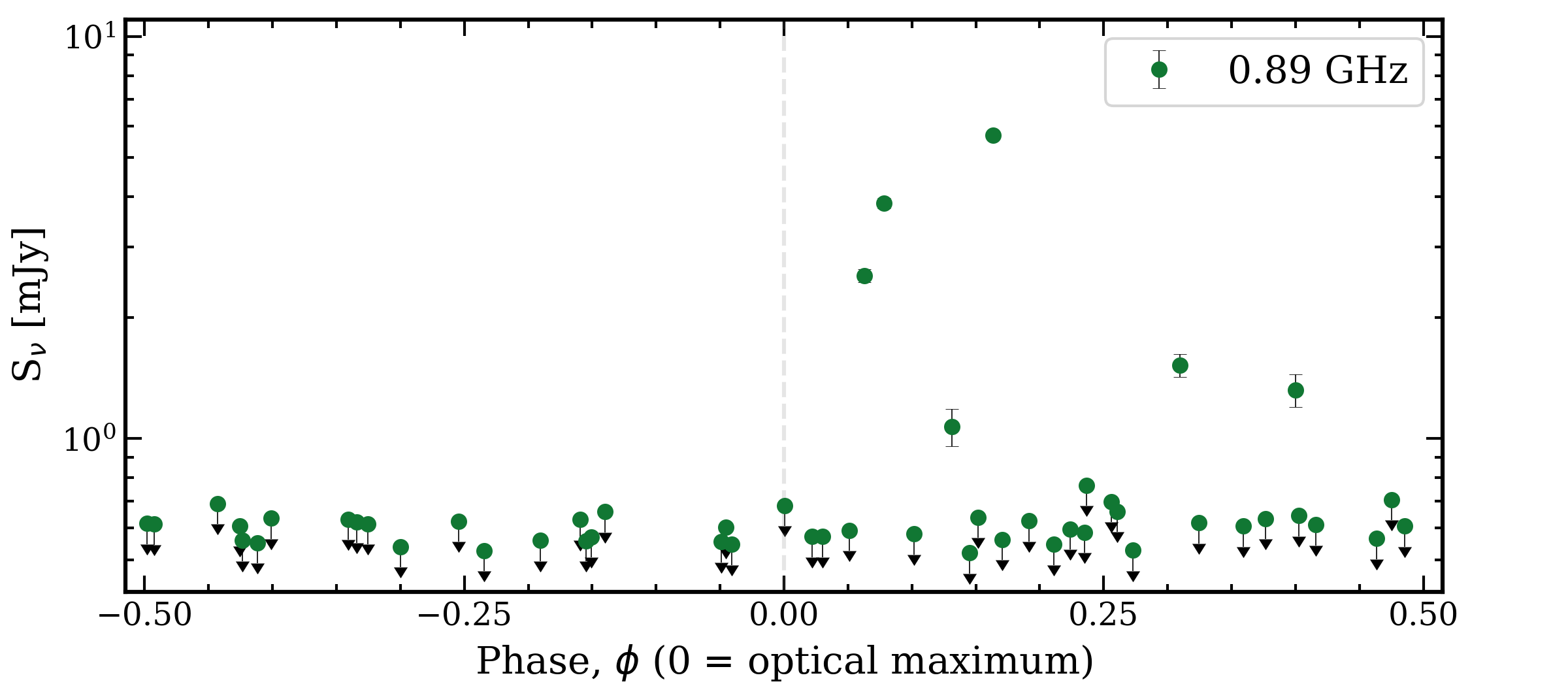}
    \label{fig:vast:phasefold}
    \vspace{-1.7em}
  \end{subfigure}
  \caption{888~MHz flux densities (top) and phase-folded light curve (bottom) from the VAST survey on ASKAP. Grey lines mark the dates of the optical maxima, arrows denote 3$\sigma_\text{rms}$ upper limits (where $\sigma_\text{rms}$ is the image root-mean-square noise), and error bars show the 1$\sigma$ statistical uncertainties.}
  \label{fig: a0535_detection}
\end{figure}

We discovered radio emission from A0538–66 using data from the Variables and Slow Transients Survey \citep[VAST;][]{Murphy2013} on the Australian Square Kilometre Array Pathfinder (ASKAP; \citealt{Hotan2021}), which comprises 36 12-m dishes in Western Australia. VAST observes 40 fields in the Galactic plane and two in the Magellanic Clouds at roughly fortnightly cadence \citep{Murphy2021PASA...38...54M_pilot}, achieving typical sensitivities of $\sim$250~$\mu$Jy beam$^{-1}$ across the full survey footprint. The data were processed with \texttt{ASKAPsoft} \citep{cornwell2011askap} to produce calibrated visibilities and full-Stokes images, after which catalogues of noise maps and total intensities were generated with \texttt{selavy} \citep{Whiting_Humphreys_2012}. The VAST pipeline \citep{adam_stewart_2024_14048598} uses these data products and source catalogues to perform source detection, association, and forced photometry, producing a set of light curves which become accessible via a web-interface and the \texttt{vast-tools} Python package \citep{adam_stewart_2025_15363128}. 

A0538–66 was detected in data from the VAST pilot Galactic surveys (2022--2024) using a machine-learning-based anomaly programme (\citealt{Andersson2025}; Andersson et al. in prep.), which identified its light curve as highly anomalous. The search used code adapted from \texttt{Astronomaly} \citep{Lochner2021}, which computes features for each light curve and applies anomaly-detection algorithms to assign an anomaly score. Sources are then presented to users via an interface that iteratively re-ranks them to reflect the required scientific goals (active learning). Independently, A0538–66 was detected with an XRB monitoring programme (Qiu et al. in prep.) which conducted a pilot census of XRB variability during the surveys.

The VAST light curve for A0538–66, spanning 2022 November 14 (MJD 59897) to 2024 November 19 (MJD 60633), is shown in Figure \ref{fig: a0535_detection}. The source was detected multiple times, notably peaking at $\sim$6~mJy in 2022 December -- a few weeks after renewed activity was reported \citep{Stubbings2022_ATel} -- and rebrightening to $\sim$4~mJy in the final 2024 November epoch. The plot for the phase-folded flux densities (lower panel) shows that all the strong radio flares occurred within the $\sim$0.2 phase following the optical maximum, likely around (and shortly after) periastron passage.  

\subsection{MeerKAT}\label{sec: meerkat}

Subsequent to its radio detection, we followed up with observations of A0538–66 through the \emph{X-KAT} programme -- the successor of \emph{ThunderKAT} \citep{thunderkat_2018} -- which conducts weekly radio monitoring of active XRBs with the MeerKAT telescope in South Africa (\citealt{jonas_2016}). The array comprises 64 13.5-m dishes, and is equipped with L-, S-, and UHF-band receivers. It has a dense core and maximum baseline of $\sim$8 km, providing excellent snapshot $\emph{uv}$ coverage. In the L-band (0.856–1.712 GHz; centered at 1.284 GHz, with a 856 MHz bandwidth), it covers a large field of view of 1.69 deg$^2$, and achieves a resolution of $\sim$5 arcsec ("). 

\subsubsection{Calibration \& Imaging}

MeerKAT's higher sensitivity compared to ASKAP allowed us to probe much lower flux densities for A0538–66. Our first L-band observation on 2025 February 4 (MJD 60710) detected an unresolved $\sim$0.2 mJy source at its reported position. We then monitored the target at L-band at roughly weekly cadence until 2025 October 12 (MJD 60960), yielding a total of 35 epochs -- one of which (ID 1742666474) was excluded from our analysis due to calibration issues. 

During each epoch, the bright and compact active galactic nucleus (AGN) J0408–6545\big/PKS 0408–65 was used as the primary (i.e., delay, bandpass, flux density), secondary (i.e., complex gain), and leakage calibrator, while the strongly polarised AGN J0521+1638\big/3C138 was used as the cross-hand phase calibrator. Each block consisted of the sequence: (J0408–6545: 5 min), (A0538–66: 15 min), (J0408–6545: 5 min), (J0521+1638: 10 min). 

The data were processed using the semi-automated routine \texttt{polkat} \citep{hughes_2025_polkat}. Using this pipeline, the visibilities were first frequency-averaged to 1024 channels. Initial flagging and reference calibration were then performed with \texttt{CASA} \citep{casa_2022} using the calibrator fields. The target data were further flagged with \texttt{tricolor} \citep{hugo_tricolour_2022}, and an initial image was produced with \texttt{WSCLEAN} \citep{offringa_wsclean_2014}. A mask generated with \texttt{breizorro} \citep{breizorro_2023} was applied for a first masked deconvolution, and the resulting model was used for direction-independent self-calibration with \texttt{QuartiCal} \citep{kenyon_2025}. A second masked deconvolution of the self-calibrated visibilities produced 16 channelised images, and a high-sensitivity multi-frequency synthesis (MFS) image. We adopted a Briggs robustness parameter \citep{briggs_1995} of 0 to optimise sensitivity while mitigating non-Gaussian features in the synthesised beam that arise at high Briggs weightings.

\begin{figure}
    \centering
    \includegraphics[width=\columnwidth]{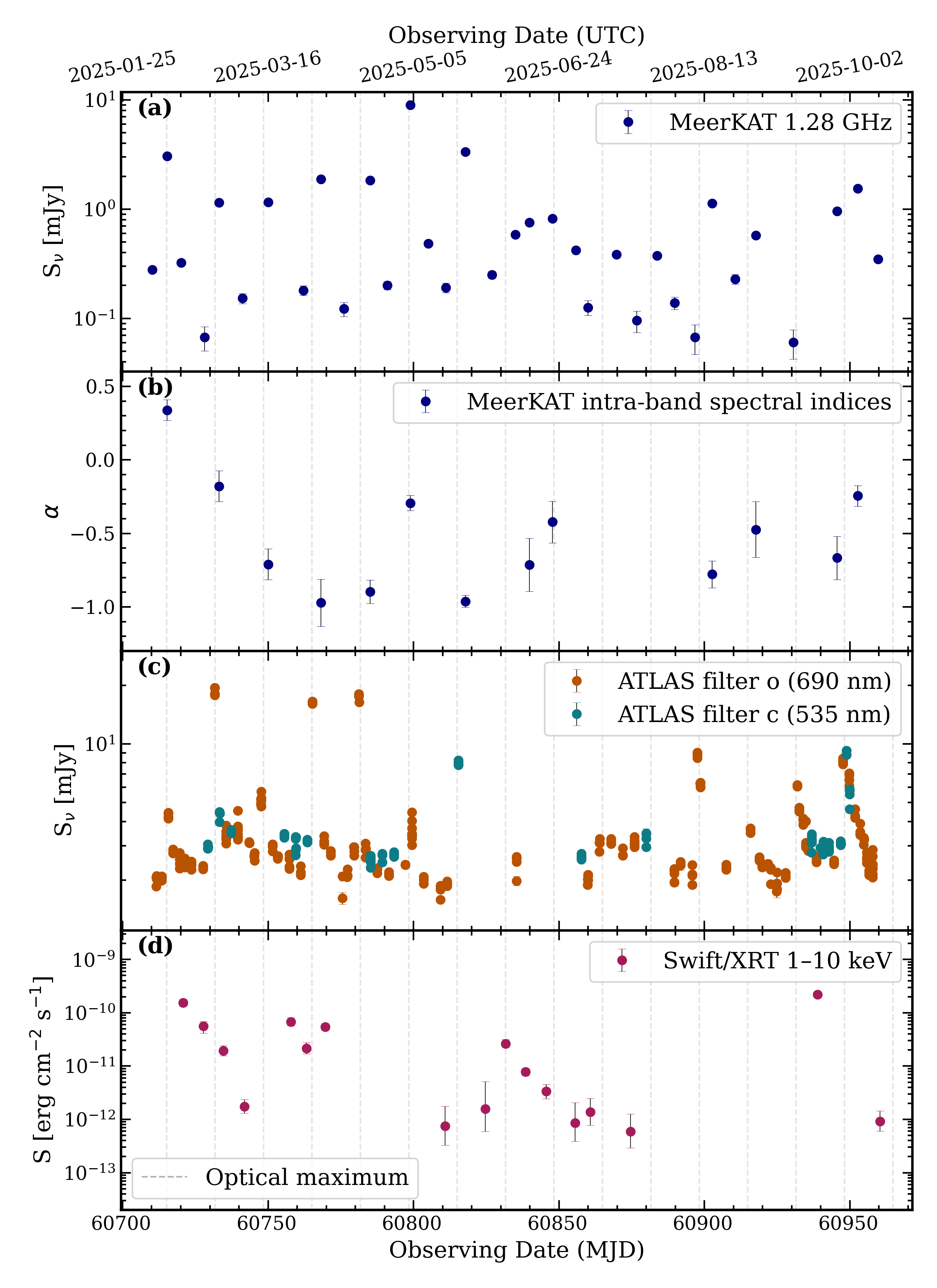} 
    \vspace{-2.5em}
    \caption{(a) 1.28 GHz MeerKAT radio core flux densities, and (b) intra-band spectral indices. (c) ATLAS filter o (560–820 nm) and c (420–650 nm) flux densities. (d) Swift/XRT unabsorbed 1–10 keV fluxes. Grey lines indicate optical maxima, and error bars show 1$\sigma$ statistical uncertainties.}
    \label{fig: all_data}
\end{figure}

\subsubsection{Analysis} 

\begin{figure}
    \centering
    \includegraphics[width=\columnwidth]{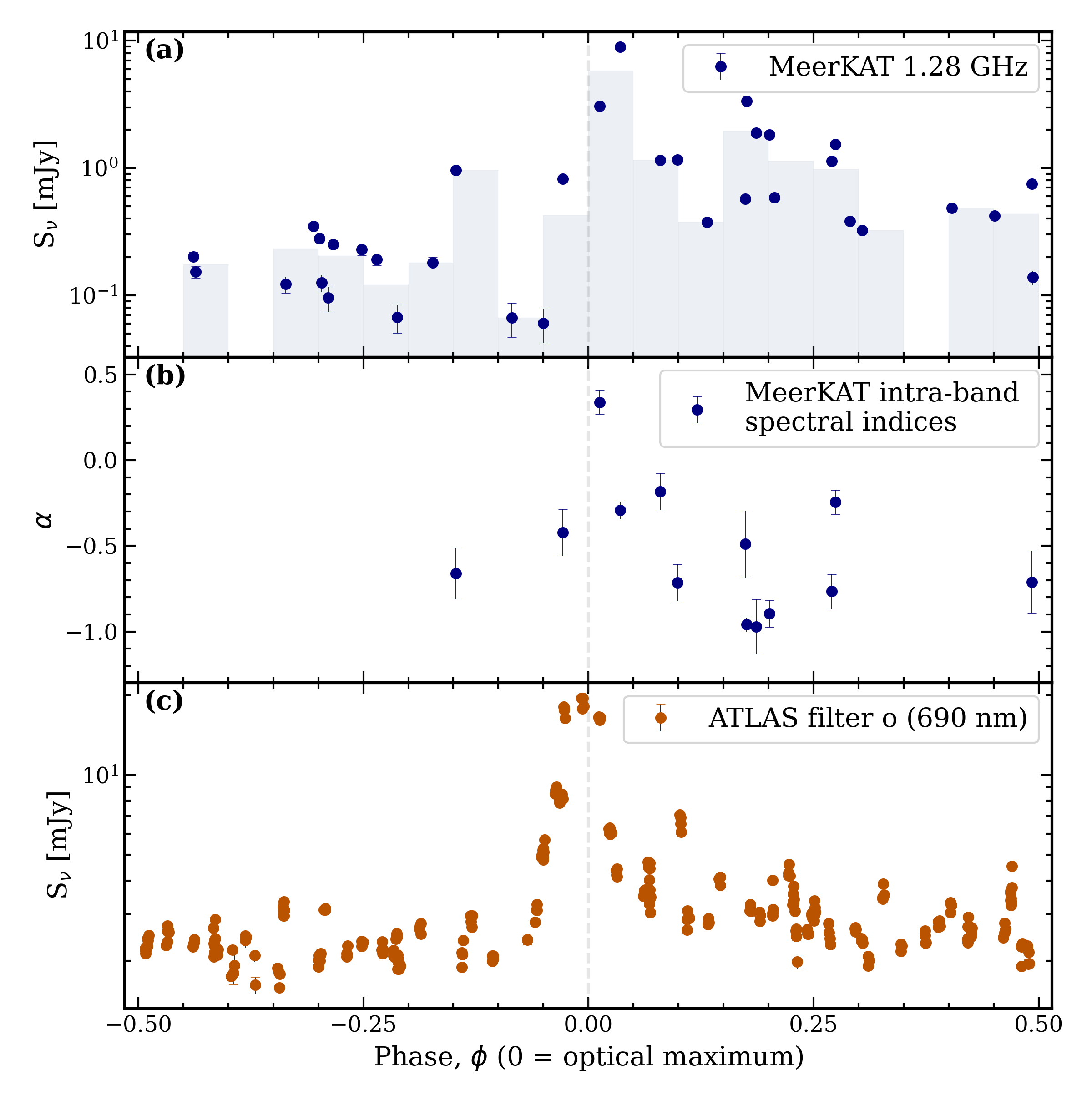} 
     \vspace{-2.5em}
    \caption[]{Phase-folded 1.28 GHz MeerKAT (a) flux densities and (b) intra-band spectral indices, and (c) ATLAS filter o (560–820 nm) data, over the date range MJDs = [60710, 60961] (covering multiple orbital cycles), where phase zero is optical maximum. The light blue shading in (a) shows average data in $\sim$0.8-day bins. Phase-folded Swift/XRT data are not shown, as they do not appear to be orbitally modulated.}
    \label{fig: a0535_folded}
\end{figure}

The root-mean-square noise ($\equiv \sigma_\text{rms}$) in each MFS image was measured in a source-free 4 arcmin$^2$ region near the source, using the \texttt{CASA} task \texttt{imstat}, yielding typical values of $\sim$20 $\mu\mathrm{Jy}$ beam$^{-1}$. The source was detected (peak flux density $>$3$\sigma_\text{rms}$) in every epoch and was unresolved. Flux densities were obtained with \texttt{imfit} by fitting an elliptical Gaussian whose shape was fixed to the synthesised beam (i.e., a point-source model), using a small region around the source. The results are shown in Figure \ref{fig: all_data}(a), with error bars indicating the 1$\sigma$ uncertainties from the fits (an additional $\sim$5--10\% systematic uncertainty is recommended). In the corresponding phase-folded plot in Figure \ref{fig: a0535_folded}(a), the light blue shading shows the average flux densities in $\sim$0.8-day bins. 

We computed intra-band spectral indices only for epochs with peak MFS source flux densities that are $\gtrsim$30 $\sigma_\text{rms}$, to avoid biases introduced when low signal-to-noise data are used \citep{heywood_2016}. For each of these epochs, the 16 sub-band flux densities ($S_\nu$) were fit as a function of frequency ($\nu$) using a power-law model ($S_\nu \propto \nu^{\alpha}$, where $\alpha$ is the spectral index). To achieve this, the data along each axis were log-transformed, and a Python implementation of \texttt{linmix} \citep{kelly_2007} was employed, which performs Markov Chain Monte Carlo simulations to fit a linear model, accounting for measurement uncertainties, upper limits in the dependent variable, and intrinsic scatter. Given that we report only the brightest epochs (which yield detections in all sub-bands), we also used a simple closed-form weighted least-squares fit, obtaining slopes (i.e., estimates for $\alpha$) that are consistent within uncertainties.

We did not detect any significant polarised emission spatially coincident with the Stokes~I source. Using the brightest epoch (2025~May~4; MJD~60809; $\text{I}\,{=}\,8.78\,{\pm}\,0.02$~mJy), we extracted Stokes~V and linear polarisation flux densities \big($\text{P}\,{=}\,\sqrt{\text{Q}^2 + \text{U}^2}$\big) by performing beam-shaped forced aperture photometry fixed at the Stokes~I position. From these flux densities, we derived a $3\sigma$ confidence interval on the circular polarisation fraction of $-0.5\% \,{<}\, m_c \,{\equiv}\, \text{V}/\text{I} \,{<}\, 0.4\%$, and a one-sided ${\approx}\,3\sigma$ upper limit on the linear polarisation fraction of $m_l\,{\equiv}\,\text{P}/\text{I}\,{<}\, 1.1\%$, following the Ricean statistical approach of \citet{Vaillancourt2006_pol_limits}. Since image-plane methods are susceptible to bandwidth depolarisation, we re-imaged the epoch with 1024 frequency channels and applied rotation measure (RM) synthesis using \texttt{RM-tools} \citep{Brentjens2005_rmsynth, cirada_rmsynth}. At this level of channelisation, we are unaffected by bandwidth depolarisation for $|\mathrm{RM}|\,{\lesssim}\,10^{4}\,\mathrm{rad\,m^{-2}}$. The resulting Faraday dispersion function shows a marginal peak at $\mathrm{RM}\,{\approx}\,1600\,\mathrm{rad\,m^{-2}}$ with a Gaussian-equivalent significance of $2.3\sigma$ \citep{Hales_2012_rmsynth_error}, and yields the same constraint on $m_l$ as the image-plane analysis. This RM is approximately an order of magnitude larger than typical pulsar values in the Magellanic Clouds ($|\mathrm{RM}|\,{\lesssim}\,100\,\mathrm{rad\,m^{-2}}$; e.g., \citealt{meerkat_mc_2022_pulsars}), which supports the interpretation that the signal is spurious. Consequently, we favour the interpretation that A0538$-$66 is weakly polarised, although whether this is intrinsic or the result of depolarisation by various factors (e.g., \citealt{pasetto_2018,hughes_2023,marscher_2021,baglio_2025,chernyakova_2024}) remains uncertain and will be investigated in future analyses of the polarisation properties across all epochs. 




We also produced an $\sim$8.5-hour deep image of the field of A0538–66 ($\sigma_\text{rms}$$\sim$10 $\mu$Jy beam$^{-1}$), but did not detect its natal supernova remnant (SNR). Indeed, the source has been associated with the LMC open cluster NGC 2034, implying an age $\gtrsim$$10^{6}$ years \citep{pakull_1981}, in which case its natal SNR (which is expected to fade on $\lesssim$$10^{5}$-year timescales; e.g. \citealt{leahy_2022}) would not be detectable.

\subsection{Swift/XRT}\label{sec: swift}

X-KAT runs in conjunction with the \emph{SwiftKAT} programme, which obtained quasi-simultaneous 0.2–10 keV observations of A0538–66 with the Neil Gehrels Swift Observatory X-ray Telescope (Swift/XRT; \citealt{gehrels_2004, burrows_swift_2005}). The spectra were extracted using the \texttt{SWIFTTOOLS} product generator \citep{evans_2007, evans_2009}, which applies the latest software and calibration files. Spectral fitting was performed with \texttt{XSPEC} \citep{arnaud_1996} over the range 0.5–10 keV, ignoring bad channels and using Cash statistics (\texttt{cstat}; \citealt{cash_1979}) after data were grouped to have at least one count per bin. Since our aim is to qualitatively track the X-ray flux evolution during our radio campaign, we adopted a simple fitting routine, reserving a more detailed analysis for future work. We extracted unabsorbed 1--10 keV fluxes following the approach of previous studies (e.g., \citealt{ducci_2019c}), adopting \texttt{tbabs*(pegpwrlw)} for low-count spectra ($<$50 counts), and the phenomenological model \texttt{tbabs*(pegpwrlw+pegpwrlw)} to capture the spectral features at both high and low energies in the higher-count spectra. For \texttt{tbabs} -- parametrised by the line-of-sight column density $N_H$ -- we used \texttt{wilm} abundances \citep{wilms_2000} and \texttt{vern} cross-sections \citep{verner_1996}. Since our spectra are mostly low-count (thus $N_H$ is poorly constrained), we fixed $N_H = 8 \times 10^{20}~\text{cm}^{-2}$, but note that varying its value across the range reported in the literature \citep{ducci_2019c,rigoselli_2025,ducci_2025} does not significantly affect our results. Uncertainties were extracted using the \texttt{error} command at the 1$\sigma$ level, and the results are shown in Figure \ref{fig: all_data}(d).

\subsection{ATLAS}\label{sec: atlas}

The Asteroid Terrestrial-impact Last Alert System (ATLAS; \citealt{tonry_2018}) consists of four telescopes and can survey the entire visible sky multiple times a day. Although it was designed to detect potentially hazardous near-Earth objects, its data are valuable for transient sources. We used ATLAS measurements of A0538–66 in the orange (o; 560–820 nm) and cyan (c; 420–650 nm) bands, generated with target-image photometry via the ATLAS server \citep{shingles_2021}. Figure \ref{fig: all_data}(c) shows the ATLAS light curve obtained during our MeerKAT campaign (MJDs 60710–60961), and Figure \ref{fig: a0535_folded}(c) presents the same data folded on the orbital period.


\section{Discussion}\label{sec: discussion}

As seen in Figure~\ref{fig: a0535_folded}(a), the radio emission of A0538–66 appears to be orbitally modulated -- with a peak occurring after the optical maximum, near periastron -- and is generally higher after the peak ($0.05 < \phi \leq 0.5$) than before ($-0.5<\phi \leq -0.05$). However, the behaviour is clearly complex, and the flux density at a particular point in time may be the superposition of various factors such as the orbital (and super-orbital) phase and accretion rate. Notably, the $\sim$1 mJy epoch at $\phi\sim-0.15$ occurred on 2025 September 28 (MJD 60581), just two days before the super-Eddington X-ray observation reported by \cite{ducci_2025_atel}, and appears brighter than data points at similar phases. Additionally, from Figure~\ref{fig: a0535_folded}(b), we see a slight tendency for the radio emission to be more optically thick ($\alpha \gtrsim -0.5$) near periastron -- although more data are needed to be certain, as we report on very few intra-band spectral indices far from periastron because of generally low signal-to-noise ratios at these phases. 

It is worth noting that the TRAPUM (TRAnsients and PUlsars with MeerKAT) Large Survey Project \citep{prayag_2024} did not detect any pulsed radio emission from A0538–66 during a 2-hour observation on 2024 March 2 (MJD 60371; $\phi \sim 0.3$), $\sim$5 days after its optical maximum (Prayag et al. in prep.). Our nearest ASKAP observation, taken $\sim$6 days later, near apastron, yielded a $3\sigma_\text{rms}$ upper limit of $\sim$0.6 mJy.

Furthermore, the optical flaring seen in Figure~\ref{fig: all_data}(c) indicates that the source is in an \emph{active} phase during our monitoring period. In particular, Figure~\ref{fig: a0535_folded}(c) shows that near periastron, an initial strong optical peak occurs, followed by additional flaring. This suggests efficient accretion near periastron (if the emission is indeed dominated by reprocessed X-rays; \citealt{ducci_2019b}). 

However, Figure~\ref{fig: all_data}(d) shows no sign of the X-ray flux being orbitally modulated (in fact, it shows large variability at the same phase), and that there is no correlation between the radio and X-ray emission. This may partly be due to our sparse sampling and lack of X-ray coverage within $\sim$1 day of periastron. Additionally, the source has previously exhibited X-ray flares far from periastron passages (see Figure 5 in \citealt{ducci_2022}), and bright outbursts can last several days (unlike the more localised optical peaks). Dense, quasi-simultaneous radio and X-ray monitoring are required to further investigate the complex behaviour of A0538–66.

\subsection{Radio Emission Mechanisms}

In this section, we consider several mechanisms that could produce radio emission in Be/XRBs, but note that the discussion is not exhaustive. 

\emph{Thermal free–free} emission from an ionised stellar wind is possible \citep{wright_1975, panagia_1975}, although it typically produces positive spectral indices, contrary to most of our observations. Moreover, the emission is classically predicted to scale with the wind mass-loss rate as $S_\nu \propto \dot{M}^{4/3}$, so  detectable L-band emission at 50 kpc likely requires $\dot{M}\gtrsim10^{-4} M_\odot~\text{yr}^{-1}$, far above the $\dot{M} \lesssim 10^{-9} M_\odot~\text{yr}^{-1}$ typical of classical Be stars \citep{krtivcka_2014}. 

A more plausible interpretation is \emph{synchrotron} emission. Assuming that the peak near periastron is due to synchrotron self-absorption (SSA; \citealt{vdL}), we can obtain \emph{order-of-magnitude} estimates of the properties of the emitting region, using the equations in \cite{fender_bright_2019} and Cowie \& Fender (in prep.) (assuming electrons with an energy index $p=2$ and Lorentz factors 3–1000). This assumption is motivated by the observation of an inverted (i.e., positive) spectral index, and similar behaviour in other flaring XRBs (e.g., \citealt{calvelo_2012}). Using our peak flux density ($\sim$$9~\text{mJy}$), the source distance, and observing frequency, we estimate a minimum energy of $\sim$$2\times10^{41}~\text{erg}$, an emitting region size of $\sim$$2\times10^{14}~\text{cm}$, and an equipartition magnetic field of $\sim$$0.2~\text{G}$. For comparison, this minimum flare energy is about two orders of magnitude higher than that of the $\sim$4 mJy flare from the NS LMXB XTE J1701–462 during an outburst that covered a similar range in X-ray luminosities (\citealt{gasealahwe_2024}; $D \approx 8.8$ kpc). Assuming that A0538–66 exhibits a radio flare every periastron (supported by our light curve), the lower limit for the power required to produce the radio emission is $\sim$$10^{35}~\text{erg~s}^{-1}\approx10^{-3}~L_\text{Edd}$. Even if the peak is not due to SSA -- but is still produced by a power-law population of synchrotron-emitting electrons (whatever the geometry) -- the minimum energy and power constraints remain valid, and the emitting region size becomes a lower limit. Bulk relativistic motion of the radio-emitting region could alter these estimates, but for a large range of parameter space, the energy remains a lower limit \citep{fender_bright_2019}. Additionally, changes in the line of sight or the velocity of bulk relativistic motion could explain the observed scatter in the light curve.

The synchrotron emission could originate from \emph{relativistic jets or ejecta}, as commonly seen in LMXBs \citep{fender_2006}. This mechanism has also been proposed for Be/XRBs -- first for Swift J0243.6+6124 \citep{van_den_eijnden_2018, van_den_eijnden_2019}, and more recently for 1A 0535+262 \citep{van_den_eijnden_2022} and LS V +44 17 \citep{van_den_eijnden_2024}. The predominantly optically thin spectral index of A0538–66 suggests emission from discrete, short-lived ejecta rather than a steady compact jet, and the jet axis could be precessing as seen in Circinus X–1 (Cir X-1; \citealt{cowie_2025}). 

Alternatively, the synchrotron emission could originate from shocks between a \emph{pulsar wind} and the Be companion's stellar wind or decretion disc, as inferred in $\gamma$-ray binaries (see \citealt{dubus_2013} for a review) such as PSR B1259–63 \citep{tavani_1997, johnston_2005}. Another possibility is that it arises due to outflows powered by a magnetospheric \emph{propeller} mechanism \citep{illarionov_1975, romanova_2005}. 

The emission mechanism that is operating depends on the relative positions of three radii: 
\begin{enumerate}[label=(\roman*),
                  leftmargin=6pt,    
                  labelsep=2pt,      
                  itemsep=0pt,
                  parsep=0pt,
                  topsep=1pt,
                  partopsep=0pt]
\item light-cylinder radius $R_{\rm lc}\simeq 4.8\times 10^{9} P_{\rm s}~\mathrm{cm} = 3.3\times10^8$ cm
\item co-rotation radius $R_{\rm co} \simeq1.68\times10^{8}\,M_{\scriptscriptstyle 1.4}^{1/3}P_{\rm s}^{2/3}\ \mathrm{cm} = 2.8\times10^7$ cm
\item magnetospheric radius $R_\text{m} \simeq 2.5\times10^{8}kM_{\scriptscriptstyle1.4}^{1/7}R_{\scriptscriptstyle6}^{10/7}B_{\scriptscriptstyle12}^{4/7}L_{\scriptscriptstyle37}^{-2/7}\mathrm{cm}$ 
\end{enumerate}
\noindent where $P_{\rm s}$ is the spin period in seconds, and the NS mass, radius, surface magnetic field, and luminosity are respectively scaled as $M = M_{\scriptscriptstyle1.4} 1.4M_\odot$, $R = R_{\scriptscriptstyle6} 10^6$ cm, $B = B_{\scriptscriptstyle12} 10^{12}$G, and $L = L_{\scriptscriptstyle37} 10^{37}\,\mathrm{erg\,s^{-1}}$ (we use $R_{\scriptscriptstyle6}=1$, $M_{\scriptscriptstyle1.4}=1$). Note that the location of $R_\text{m}$ is quite uncertain; we use \citealt{tsygankov_2017} equation 2, and assume $k=0.5$ for disc accretion, although \citealt{ducci_2019c} suggested that A0538–66 may sometimes be in a regime of spherical accretion. The source is in a direct-accretion regime when $R_{\rm m} < R_{\rm co}$, a propeller outflow occurs when $R_{\rm lc} > R_{\rm m} > R_{\rm co}$, and an intra-binary shock can form when $R_{\rm m} > R_{\rm lc}$ (respectively cases A/B, C and D in the simulations by \citealt{2017ApJ...851L..34P}). 

\citet{ducci_2025} considered two scenarios to explain the presence of X-ray pulsations in A0538–66 at $L_{X_0} \sim 8\times10^{36}$ erg s$^{-1}$. In the \emph{first scenario}, the NS is directly accreting at $L_{X_0}$, implying $B<7\times10^{10}$ G, which would be unusually low for a NS HMXB (see discussion in \citealt{ducci_2025}). In this case, the source would be in either the propeller or direct-accretion regimes during all epochs in our campaign, so the radio emission likely arises from a propeller outflow or jet ejecta. 

In the \emph{second scenario}, the NS resides in a propeller phase at $L_{X_0}$ -- so can have a stronger magnetic field of  $B<5\times10^{12}$ G (using $R_{\rm lc} > R_{\rm m}$) -- and pulsations arise when gas intermittently overcomes the magnetospheric barrier and accretes onto the NS poles \citep[in between cases B and C in][]{2017ApJ...851L..34P}. In this case, the source could transition between the propeller and pulsar-wind shock regimes at some $L_X < L_{X_0}$ that depends on the exact magnetic field strength (i.e., a `flip-flop' scenario, as proposed for the $\gamma$-ray binary LS I +61$^\circ$303; \citealt{torres_2012}). We note that for $B< 5\times10^{11}$ G, this transition would fall below the lowest X-ray flux seen in our Swift/XRT monitoring, which would again suggest that all radio emission originates in an accretion-flow-dominated state. It is unclear whether two different radio emission mechanisms alternate in A0538--66, as we do not observe a clear dichotomy in the radio properties (e.g., luminosity, spectral index) above and below a certain $L_X$. If a subset of the radio observations (those at low $L_X$) originate from a shock between the pulsar wind and the Be-star's disc, the field $B<5\times10^{12}$ G implies a pulsar spin-down power of $\dot{E} \lesssim 10^{37}$ erg s$^{-1}$, which gives a ratio of radio luminosity to $\dot{E}$ that is comparable to those in two systems with intra-binary shocks and known $\dot{E}$, namely PSR B1259--63 (\citealt{johnston_1992, johnston_1996, dubus_2013, chernyakova_2024}) and PSR J2032+4127 (\citealt{lyne_2015, ng_2019, camilo_2009, ho_2017}). These systems have much wider orbits than A0538--66 (respectively $P_\text{orb} \approx 3.4$ years and $\approx 45-50$ years); thus, far from periastron, the companion-wind density at the NS is low enough for radio pulsations to be active and detectable -- a situation not observed to date in A0538–66. 



\subsection{Broader Context}

\begin{figure}
    \centering
    \includegraphics[width=\columnwidth]{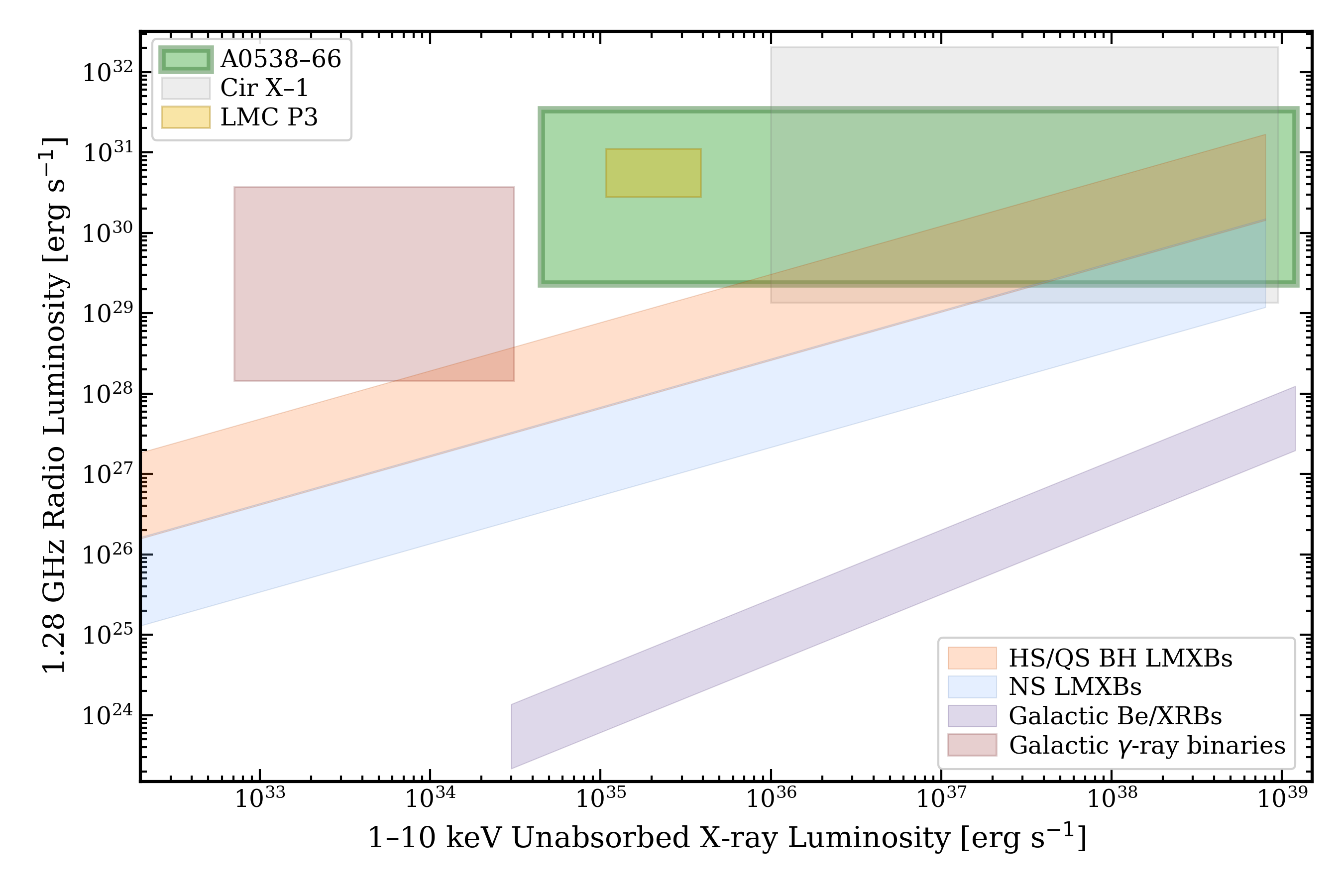} 
    \vspace{-2.5em}
    \caption{A schematic showing regions in radio:X-ray parameter space occupied by different types of systems (see the text for caveats). Orange and blue respectively indicate the hard/quiescent-state black hole and neutron star low-mass X-ray binaries. Purple and red show subsets of the Galactic Be/X-ray binary and $\gamma$-ray binary classes, respectively, while yellow indicates the extra-Galactic $\gamma$-ray binary LMC P3. A0538–66 and Cir X–1 are shown in green and grey, respectively.}
    \label{fig: lrlx_comparison}
\end{figure}

Figure \ref{fig: lrlx_comparison} shows a schematic comparison of quasi-simultaneous unabsorbed 1–10 keV X-ray luminosities and 1.28 GHz radio luminosities scaled by frequency (i.e. $L_X$ and $L_R = \nu L_\nu$), for X-ray and $\gamma$-ray binaries. The diagram is deliberately simplified and, given sensitivity limits and sparse sampling, may not capture the full parameter space of each population -- but it provides a useful high-level comparison of A0538–66 with other similar systems. The green shading indicates the $L_R$:$L_X$ parameter space occupied by A0538–66, though $L_X$ may extend down to even lower values in quiescence \citep{kretschmar_2004}. Orange and blue show the regions corresponding to the canonical hard/quiescent BH LMXB track and to NS LMXB detections, respectively (e.g., \citealt{gallo_2018}). In reality, these regions overlap: many BH LMXBs extend to lower $L_R$ (e.g., \citealt{hughes_2025}), NS LMXBs exhibit diverse behaviour (e.g., \citealt{tetarenko_2016b, tudor_2017}). Additionally, many NS observations are upper limits, so this region likely extends to even lower $L_R$ than currently detectable. 

Using observations of 1A 0535+262, SAX J2103.5+4545, an outburst of Swift J0243.6+6124, and a giant outburst of GRO J1008--57, \cite{van_den_eijnden_2022} proposed that \emph{Galactic Be/XRBs} display correlated radio and X-ray emission -- consistent with radio emission arising from accretion-driven compact jets -- contrary to our (sparse) A0538–66 observations. These authors further suggested that this group (shown in purple in Figure \ref{fig: lrlx_comparison}) has a lower $L_R/L_X$ normalisation than NS LMXBs. Thus, regardless of whether the radio emission is jet-driven or has another origin (e.g., \citealt{chatzis_2022}), Be/XRBs at comparable luminosities would not be detectable at the LMC distance -- highlighting how luminous A0538–66 is. An investigation into whether this difference may be related to A0538–66's significantly shorter spin period compared to other Be/XRBs (see \citealt{rigoselli_2025} Figure 7) is worthwhile.  

As first noted by \cite{murdin_1981}, A0538–66 resembles the NS LMXB \emph{Cir X–1} in terms of its $\sim$16.6-day orbital period (\citealt{kaluzienski_1976}) and X-ray light curve, although Cir X–1 is far younger ($<$4600 years; \citealt{heinz_2013}). Cir X–1 typically shows radio flux densities of the order $\sim$10 mJy which (at $D \approx 9.4$ kpc; \citealt{heinz_2015}) corresponds to luminosities that are comparable to A0538–66. However, it occasionally flares to $\gtrsim$1 Jy \citep{armstrong_2013}, about an order of magnitude more luminous than A0538–66's peak (observed to date). In grey in Figure \ref{fig: lrlx_comparison}, we plot a representative orbitally-averaged range of $L_X$ for Cir X–1; while it has exhibited lower $L_X$ dips near periastron, these are likely due to absorption rather than a decrease in the accretion rate \citep{dai_2012, schulz_2020}. Cir X–1 has long been regarded as an extreme outlier among NS XRBs, so our finding that A0538–66 resides in a similar region of $L_R$:$L_X$ space is particularly intriguing.

In red in Figure \ref{fig: lrlx_comparison}, we indicate the approximate location of some of the most luminous \emph{Galactic $\gamma$-ray binaries} -- 1FGL J1018.6–5856 (1FGL J1018; \citealt{van_soelen_2022}; $D \approx 4.4$ kpc), LS I +61$^\circ$303 (\citealt{esposito_2007, massi_2009, zimmermann_2015, massi_2016,massi_2017}; $D \approx 2.6$ kpc), LS 5039 (\citealt{bosch_ramon_2005,marcote_2015,volkov_2021}; $D \approx 2.0$ kpc), PSR B1259–63 (\citealt{chernyakova_2009}; $D \approx 2.4$ kpc), and 4FGL J1405.1-6119 (\citealt{corbet_2019}; $D \approx 7.7$ kpc) -- but note that there may be fainter sources within this group (e.g., PSR J2032+4127; \citealt{ho_2017,ng_2019,chernyakova_2020}; $D \approx 1.4$ kpc). Interestingly, 1FGL J1018 also has $P_{\rm orb}\approx 16.6$ days \citep{van_soelen_2022}; however, its most luminous L-band MeerKAT epoch (to date) is comparable to the least luminous A0538–66 epochs (Mathiba et al. in prep.). In addition, the yellow region in Figure \ref{fig: lrlx_comparison} shows the approximate parameter space occupied by the only confirmed \emph{extra-Galactic $\gamma$-ray binary}, LMC P3 \citep{corbet_2016}, whose L-band MeerKAT flux density peaks at slightly lower but comparable values to A0538–66 (namely $\lesssim 3$ mJy; Mathiba et al. in prep.). 

Clearly, A0538–66 straddles the region in $L_R$:$L_X$ parameter space between $\gamma$-ray binaries and outlier NS LMXB systems like Cir X-1. 


\section{Conclusions}

A0538–66 is a Be/XRB in the LMC that exhibits a range of unusual properties. We detected the source at radio wavelengths for the first time in 0.89 GHz ASKAP data. We subsequently conducted a weekly 1.28 GHz MeerKAT monitoring campaign, during which it was consistently detected and reached a peak flux density of $\sim$9 mJy. In addition, we found tentative evidence that the radio emission is orbitally modulated. In upcoming work, we will present the results from a high-cadence multi-wavelength campaign aimed at understanding the origins of emission in this peculiar system.

\section*{Acknowledgements}

JCM acknowledges financial support from the Rhodes Trust at the University of Oxford. RF acknowledges support from the European Research Council (ERC) Synergy Grant `Blackholistic', UK Research and Innovation (UKRI), and the Hintze Family Charitable Foundation. JvdE was supported by funding from the European Union's Horizon Europe research and innovation programme under the Marie Skłodowska-Curie grant agreement No 101148693 (MeerSHOCKS). DK was supported by NSF grant AST-2511757. Parts of this research were conducted by the Australian Research Council Centre of Excellence for Gravitational Wave Discovery (OzGrav), project number CE230100016. LD acknowledges funding from the Deutsche Forschungsgemeinschaft (DFG, German Research Foundation) - Projektnummer 549824807. PAC acknowledges the Leverhulme Trust for an Emeritus Fellowship.

The MeerKAT telescope is operated by the South African Radio Astronomy Observatory, which is a facility of the National Research Foundation, an agency of the Department of Science and Innovation. The authors would like to thank Lilia Tremou, Andrew Hughes, Francesco Carotenuto, and Payaswini Saikia for scheduling the MeerKAT observations. We acknowledge the use of the Inter-University Institute for Data Intensive Astronomy (IDIA) data intensive research cloud for data processing. IDIA is a South African university partnership involving the University of Cape Town, the University of Pretoria, and the University of the Western Cape. 

This scientific work uses data obtained from Inyarrimanha Ilgari Bundara, the CSIRO Murchison Radio-astronomy Observatory. We acknowledge the Wajarri Yamaji People as the Traditional Owners and native title holders of the Observatory site. CSIRO’s ASKAP radio telescope is part of the Australia Telescope National Facility (\url{https://ror.org/05qajvd42}). Operation of ASKAP is funded by the Australian Government with support from the National Collaborative Research Infrastructure Strategy. ASKAP uses the resources of the Pawsey Supercomputing Research Centre. Establishment of ASKAP, Inyarrimanha Ilgari Bundara, the CSIRO Murchison Radio-astronomy Observatory, and the Pawsey Supercomputing Research Centre are initiatives of the Australian Government, with support from the Government of Western Australia and the Science and Industry Endowment Fund.

This research was supported by the Sydney Informatics Hub (SIH), a core research facility at the University of Sydney. This work was also supported by software support resources awarded under the Astronomy Data and Computing Services (ADACS) Merit Allocation Program. ADACS is funded from the Astronomy National Collaborative Research Infrastructure Strategy (NCRIS) allocation provided by the Australian Government and managed by Astronomy Australia Limited (AAL).

\section*{Data Availability and Software}

The code used to generate the plots in this paper is available at: \url{https://github.com/JustineCrook/A0538-66_analysis}; any updates to the radio and X-ray results will be shared here. Data from MeerKAT are available through the SARAO data archive (Proposal ID: SCI-20230907-RF-01): \url{https://archive.sarao.ac.za/}. Data from ATLAS are available at: \url{https://fallingstar-data.com/forcedphot/}. Data from the Swift/XRT are publicly available through the Swift archive: \url{https://www.swift.ac.uk/swift_portal}.

For the MeerKAT data reduction, we made use of \texttt{polkat}: \url{https://github.com/AKHughes1994/polkat}. When fitting the MeerKAT intra-band spectral indices, we used the Python implementation of \texttt{linmix} from: \url{https://github.com/jmeyers314/linmix}.



\bibliographystyle{mnras}
\bibliography{references} 





\bsp	
\label{lastpage}
\end{document}